\documentclass[conference,a4paper]{IEEEtran}
\IEEEoverridecommandlockouts
\usepackage{cite}
\usepackage{amsmath,amssymb,amsfonts}
\usepackage{algorithmic}
\usepackage{subfigure}
\usepackage{graphicx}
\usepackage{textcomp}
\usepackage{xcolor}
\usepackage{balance}
\usepackage[inkscapelatex=false]{svg}
\usepackage{algorithm}
\usepackage[font=footnotesize]{subcaption}
\usepackage[font=footnotesize]{caption}

\usepackage[left=1.42cm,right=1.41cm,top=1.79cm,bottom=4.35cm]{geometry} 
\def\BibTeX{{\rm B\kern-.05em{\sc i\kern-.025em b}\kern-.08em
    T\kern-.1667em\lower.7ex\hbox{E}\kern-.125emX}}
\begin{document}

\newcommand{\cel}[1]{{\color{cyan}[Cel: #1]}}
\newcommand{\zc}[1]{{\color{blue}[Zach: #1]}}
\newcommand{\robb}[1]{{\color{blue}[Robbert: #1]}}

\title{Near-Field Spatial non-Stationary Channel Estimation: Visibility-Region-HMM-Aided Polar-Domain Simultaneous OMP}
\author{Thibaut Ceulemans, Cel Thys, Robbert Beerten, Zhuangzhuang Cui, Sofie Pollin\\
Department of Electrical Engineering (ESAT), KU Leuven, 3001 Leuven, Belgium\\
Email: \texttt{thibaut.ceulemans@student.kuleuven.be}}

\maketitle

\begin{abstract}
This work focuses on channel estimation in extremely large aperture array (ELAA) systems, where near-field propagation and spatial non-stationarity introduce complexities that hinder the effectiveness of traditional estimation techniques. A physics-based hybrid channel model is developed, incorporating non-binary visibility region (VR) masks to simulate diffraction-induced power variations across the antenna array. To address the estimation challenges posed by these channel conditions, a novel algorithm is proposed: Visibility-Region-HMM-Aided Polar-Domain Simultaneous Orthogonal Matching Pursuit (VR-HMM-P-SOMP). The method extends a greedy sparse recovery framework by integrating VR estimation through a hidden Markov model (HMM), using a novel emission formulation and Viterbi decoding. This allows the algorithm to adaptively mask steering vectors and account for spatial non-stationarity at the antenna level. Simulation results demonstrate that the proposed method enhances estimation accuracy compared to existing techniques, particularly in low-SNR and sparse scenarios, while maintaining a low computational complexity. The algorithm presents robustness across a range of design parameters and channel conditions, offering a practical solution for ELAA systems.
\end{abstract}


\section{Introduction}
The development of sixth-generation (6G) wireless networks will introduce new performance requirements and application scenarios, including high-precision localization, ultra-reliable low-latency communication, and high-capacity connectivity~\cite{ITU-R_report_3-2516-0, wang_road_2023}. To meet these demands, one of the key enabling technologies under consideration is the use of an extremely large aperture array (ELAA)~\cite{lu_tutorial_2024}, which involves deploying antenna arrays with hundreds of elements.

Unlike conventional antenna systems, ELAAs operate predominantly in the near-field region due to their large physical aperture. In this regime, electromagnetic wavefronts can no longer be approximated as planar, and instead require spherical modeling~\cite{zhou_spherical_2015, liu_near-field_2024}. Additionally, the large array size leads to spatial non-stationarity (SnS), where different parts of the array observe different subsets of propagation paths~\cite{martinez2014towards, de2020non}. These characteristics fundamentally change the behavior of the wireless channel relative to traditional models and introduce new challenges for signal processing.

Accurate channel estimation is critical in ELAA systems, as it underpins key functionalities such as beamforming, spatial multiplexing, and user localization~\cite{Rusek_Scaling_2013}. However, the near-field and spatially non-stationary nature of the channel complicates the estimation process. Traditional methods, which assume far-field propagation and spatial stationarity, are not well-suited for this setting.

Due to increased interest at high frequencies, particularly in the millimeter-wave (mmWave) bands, signals propagate through only a limited number of dominant paths~\cite{Sloane_2023-measurement, he2021wireless, Zhang_2024-DeterministicRay, Liu2024ChannelSparsity}. Combined with the high spatial resolution of ELAA systems, this leads to a sparse channel structure: the number of dominant propagation paths is typically much smaller than the number of paths that can be resolved by the antenna array. This sparsity opens the door for compressed sensing and sparse recovery techniques, which can reduce pilot overhead and computational complexity~\cite{bajwa_compressed_2010, berger_application_2010, mansoor_massive-mimo_2017}. Designing channel estimation algorithms that can effectively exploit sparsity while accounting for near-field effects and spatial non-stationarity remains an open and important problem.

\subsection{Prior Works}
Exploiting the channel sparsity in extremely large aperture array systems has been widely studied. Compressed sensing and sparse recovery techniques have received significant attention in this context. In compressed sensing, the channel is transformed into a domain where it can be represented sparsely, i.e., using only a few non-zero coefficients. Sparse recovery algorithms are then used to reconstruct this sparse representation from noisy measurements.

Several approaches have been proposed for sparse recovery in channel estimation. Convex relaxation techniques, such as the Least Absolute Shrinkage and Selection Operator (LASSO), have been explored in works such as~\cite{Liu_2016_joint-burst, schniter2014channel,destino2015leveraging}. Other authors adopt non-convex optimization techniques to perform sparse channel recovery by directly estimating the underlying parameter distribution of the wireless channel~\cite{xu_joint_2024, yu2023channel,chen2022hierarchical,cheng2019adaptive} .

Among sparse recovery methods, greedy algorithms offer a computationally efficient alternative by iteratively constructing the sparse solution. At each step, these algorithms select the component that provides the largest local improvement to the approximation~\cite{tropp_greed_2004}. The work in~\cite{cui_channel_2022} employs the Simultaneous Orthogonal Matching Pursuit (SOMP) framework, which extends the classic OMP by jointly estimating the sparse channel representations across multiple subcarriers, thereby improving estimation accuracy. 

Traditionally, angular-domain steering vectors based on the plane wave assumption were used to define the sparse basis. However, in near-field scenarios, where the plane wave assumption breaks down, the channel can no longer be sparsely represented in the angular domain. To address this issue, polar-domain transformations have been introduced, leveraging near-field steering vectors instead of far-field approximations~\cite{cui_channel_2022}. This work aims to propose a novel method to construct a sparse basis tailored for near-field channels.

In addition to near-field propagation, spatial non-stationarity further complicates channel estimation. SnS is commonly modeled using visibility regions (VRs), where a visibility region denotes the subset of antennas over which a given propagation path is observable. Prior channel estimation research has predominantly modeled VRs using binary masks, where each element is either 0 or 1. However, the authors in~\cite{yuan_spatial_2023} have demonstrated that incorporating non-binary VR masks is essential for capturing diffraction-induced power variations across the antenna array, thereby leading to a more physically accurate channel model.

To mitigate the effects of SnS in channel estimation, several approaches have been proposed. One much visited method is subarray-wise estimation~\cite{han_channel_2020, chen_non-stationary_2024}, which partitions the antenna array into smaller subarrays, assuming that for a subarray spatial stationarity can be assumed. Subarray-based approaches introduce a trade-off: smaller sub-arrays mean finer SnS characterisation, but at the expense of reduced channel estimation accuracy within each individual subarray due to the smaller number of antennas available for estimation~\cite{chen_non-stationary_2024}.

Recently in~\cite{tang_joint_2024}, a two-stage visibility region detection and channel estimation framework (TS-VRCE) was introduced. In the first stage, a message-passing algorithm estimates the visibility region probabilities, which are then used in the second stage to modify an angular-domain orthogonal matching pursuit (OMP) algorithm for channel estimation. While TS-VRCE successfully incorporates VR awareness, it assumes a shared VR mask across all paths, which oversimplifies spatial non-stationarity~\cite{yuan_spatial_2023}. Similarly, the Alternating MAP framework in~\cite{xu_joint_2024} uses a Bayesian inference approach, alternating between VR estimation, sparse channel recovery, and adaptive grid refinement. However, it suffers from high computational complexity and requires extensive iterations for convergence. These limitations highlight the need for a low-complexity yet robust channel estimation approach capable of dynamically integrating SnS at the antenna level.

\subsection{Contributions}
This work proposes the \emph{Visibility-Region-HMM-Aided Polar-Domain
Simultaneous Orthogonal Matching Pursuit} (VR-HMM-P-SOMP), a novel OMP-based channel estimation algorithm that addresses the key challenges posed by near-field propagation and spatial non-stationarity in extremely large aperture array systems. The proposed method leverages the inherent sparsity of the wireless channel while maintaining low computational complexity, making it suitable for practical implementation. The main contributions and novelties are summarized as follows:

\begin{itemize}
    \item \textbf{Non-binary visibility region modeling:} The channel model incorporates non-binary VR masks, thereby enabling the inclusion of diffraction-based SnS characterization in the simulation-based performance evaluation.

    \item \textbf{Adaptation of the Polar-domain SOMP framework to spatial non-stationarity:} The Polar-domain Simultaneous Orthogonal Matching Pursuit (P-SOMP) algorithm is extended to support SnS of channel characteristics. This adaptation introduces two main innovations:
    \begin{itemize}
         \item \textit{Dynamic VR-masked steering vectors:} Steering vectors are dynamically masked based on the estimated VR, and this masking is interleaved directly within the OMP iteration loop.
        
        \item \textit{HMM-based VR estimation:} A hidden Markov model (HMM) is employed to estimate the VR at antenna-level for each propagation path. A novel emission formulation is introduced, and the most likely VR mask is efficiently inferred using Viterbi decoding. 
    \end{itemize}    
\end{itemize}

\subsection{Organization and Notation}
\textit{Organization:} The remainder of this paper is organized as follows: Section~\ref{sec: system model} introduces the scenario and the channel model. This model helps to better understand the characteristics of the ELAA channel and serves as the foundation for simulating the performance of various channel estimation algorithms. In Section~\ref{sec: proposed algorithm}, the proposed VR-HMM-P-SOMP algorithm is presented and discussed. Section~\ref{sec: simulation results} provides simulation results that assess the accuracy of the proposed algorithm in comparison with other benchmark greedy sparse recovery algorithms.

\textit{Notation:} Vectors and matrices are denoted by lowercase and uppercase boldface letters, respectively: $\mathbf{x}$, $\mathbf{X}$; $\mathbf{x}_{[i]}$ denotes the $i$\textsuperscript{th} element of vector $\mathbf{x}$; $\mathbf{X}_{[i,j]}$ denotes the $(i,j)$\textsuperscript{th} element of $\mathbf{X}$; $\mathbf{X}_{[:,j]}$ and $\mathbf{X}_{[i,:]}$ refer to the $j$\textsuperscript{th} column and the $i$\textsuperscript{th} row, respectively. Subscripts without square brackets are used to distinguish parameters or indicate loop variables, e.g.,  $x_k$ for $k\in \{1, \dots, K \}$; $(\cdot)^\mathcal{P}$ indicates a polar domain representation; $(\cdot)^T$, $(\cdot)^H$, and $(\cdot)^\dagger$ represent the transpose, Hermitian (conjugate transpose), and pseudo-inverse, respectively. The absolute value is denoted by $|\cdot|$, and $||\cdot||_n$ represents the $L_n$-norm. The Hadamard (element-wise) product is written as $\odot$. $\mathcal{C}\mathcal{N}(\mu, \sigma^2)$ is the complex Gaussian distribution with mean $\mu$ and variance $\sigma^2$, and $\mathcal{U}(a, b)$ is the uniform distribution over the interval $[a, b]$.

\section{System Model}
\label{sec: system model}
\subsection{Scenario}
The scenario considered in this paper consists of a base station (BS) equipped with $N$ antennas arranged in a uniform linear array (ULA) configuration, and $K$ single-antenna user devices communicating with the BS. To perform channel estimation, each user transmits mutually orthogonal pilot sequences of length $T$ across $M$ subcarriers simultaneously. The user equipment is assumed to be located in the near field of the BS.

When a user device sends a pilot sequence, the signal reaches the base station either through a line-of-sight (LOS) path or via scatterers located in the near field (non-line-of-sight (NLOS)). Electromagnetic propagation over the wireless channel between a user and the BS occurs via a select number of dominant paths. The number of dominant propagation paths $L$ is much smaller than the number of antennas at the BS ($L \ll N$). In addition to sparsity and near-field effects, the channel also exhibits spatial non-stationarity; different antennas may experience different visibility of the propagation paths due to phenomena such as blockage and partial reflections.

Throughout this paper, a fully digital array setup is assumed; the received signal at each antenna element is abstracted as a complex baseband symbol, capturing both the amplitude and phase of the signal on a given subcarrier after RF-chain processing, including downconversion and demodulation.

\subsection{Channel Model}
The channel model for the described scenario is based on the works of~\cite{sherman_properties_1962, cui_channel_2022, yuan_spatial_2023}. Channel estimation is performed on the received pilot symbols $\mathbf{y}_{m,t}$, as defined in Eqn.~\ref{eq: received pilot}. Here, $\mathbf{h}_m$ denotes the channel vector for subcarrier $m$, which will be further modeled. The variable $x_{m,t}$ represents the pilot symbol transmitted on subcarrier $m$ at time slot $t$, and $\mathbf{n}_{m,t}$ denotes the additive complex Gaussian noise, assumed to follow the distribution $\mathcal{CN}(\mathbf{0}, \sigma^2\mathbf{I})$. Note that every element in the vector corresponds to an antenna in the array.
\begin{equation}
\mathbf{y}_{m,t} = \mathbf{h}_{m} x_{m,t} + \mathbf{n}_{m,t} \quad \textnormal{with} \quad t = 1,...,T.
\label{eq: received pilot}
\end{equation} 

Since the pilot sequences transmitted by different users are mutually orthogonal, and without loss of generality, we consider an arbitrary user whose pilot symbols are all set to~1. This simplifies the expression to:
\begin{equation}
\mathbf{y}_{m,t} = \mathbf{h}_m + \mathbf{n}_{m,t}.
\label{eq: received pilot simple}
\end{equation} 

The channel vector $\textbf{h}_m$ remains constant during the pilot sequence of length $T$. The objective of the channel estimation algorithms is to accurately recover $\mathbf{h}_{m}$ from the received pilot symbols. The channel vector $\mathbf{h}_m$ is modeled as:
\begin{equation}
\mathbf{h}_m = \sqrt{\frac{N}{L}} \sum_{l=1}^{L} g_l e^{-j k_m r_l} \mathbf{b}(\theta_l, r_l) \odot \mathbf{s}_{m,l}
\label{eq: hm with SnS}
\end{equation} 
where each component of the model is described in more detail down below.

Firstly, each expression in the summation term $\sum(\cdot)$ represents the contribution of a specific dominant propagation path indexed by $l$. This contribution is constructed as follows: $\mathbf{b}(\theta_l, r_l)$ denotes the near-field steering vector, which characterizes the relative phase shifts across the antenna elements due to a path from a near-field scatterer. The steering vector is defined as~\cite{cui_channel_2022}:

\begin{equation}
\mathbf{b}(\theta_l, r_l) = \frac{1}{\sqrt{N}} 
\left [ 
e^{-j k_c (r_l^{(0)} - r_l)}, \dots, e^{-j k_c (r_l^{(N-1)} - r_l)}
\right ]^T,
\label{eq: b()}
\end{equation} 

where $k_c = \frac{2\pi f_c}{c}$ is the wavenumber corresponding to the carrier frequency $f_c$, with $c$ denoting the speed of light. (Note that $k_m$ is the wavenumber for subcarrier $m$.) The parameter $r_l$ is the distance from the scatterer to the center of the antenna array, while $r_l^{(n)}$ denotes the distance from the scatterer to the $n$\textsuperscript{th} antenna element. Due to the geometry of the scenario, the steering vector $\mathbf{b}(\theta_l, r_l)$ depends on both the angle of arrival $\theta_l$ and the distance $r_l$ of the propagation path relative to the array center.

The term $e^{-j k_m r_l}$ represents the phase shift due to the used subcarrier and $g_l$ is the complex path gain associated with scatterer $l$, capturing both attenuation and phase effects.

\begin{figure}
    \centering
    \includesvg[width=0.90\linewidth]{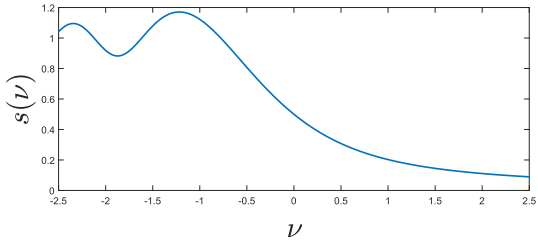}
    \caption{Diffraction gain}
    \label{fig: fresnel gain}
\end{figure}

The vector $\mathbf{s}$ is the VR mask, which models spatial non-stationarity by element-wise multiplying with the steering vector. The authors in~\cite{yuan_spatial_2023} identify three key cases that should be considered in the design of VR masks:

\begin{itemize}
    \item \textbf{Spatially stationary case:} The VR mask is set to all ones ($\mathbf{s} = \mathbf{1}$), implying that every antenna lies within the visibility region.
    
    \item \textbf{Binary case:} Due to blockage or partial reflections, only a subset of antennas are within the VR. In this case, the mask consists of ones within the VR boundaries and zeros outside. The boundaries are randomly sampled from a uniform distribution over the array length, i.e., $\mathcal{U}(1, N)$.
    
    \item \textbf{Non-binary case:} In scenarios with diffraction, the VR cannot be modeled as strictly binary. To capture diffraction-induced power variation across the antenna array, a knife-edge model is used. The mask values depend on the Fresnel diffraction gain pattern caused by an angular deviation $\theta_\text{edge}$ of the diffraction edge relative to the line-of-sight. The variation is described using the Fresnel integral parameter $\nu$~\cite{ITU_r_p.526}, defined as:
    \begin{equation}
        \nu = \theta_\text{edge} \sqrt{\frac{2}{\lambda_m \left( \frac{1}{d_1} + \frac{1}{d_2} \right)}}
        \label{eq: nu}
    \end{equation}
    where $\lambda_m$ is the subcarrier wavelength and $d_1 = d_2 = r_l$ for simplicity. During channel generation, an angular deviation $\theta_\text{edge}$ corresponding to the center of the array is drawn from a uniform distribution $\mathcal{U}(-\theta_\text{max}, \theta_\text{max})$. Based on the array geometry, the corresponding deviation $\theta_\text{edge}^{(n)}$ for each antenna element is then computed. With $\theta_\text{edge}^{(n)}$ known, the Fresnel diffraction parameter $\nu^{(n)}$ is calculated using Eqn.~\ref{eq: nu}. The diffraction gain for each antenna is subsequently obtained using the knife-edge diffraction model specified by the ITU in~\cite[Eqn.~30]{ITU_r_p.526}. These gain values, illustrated in Fig.~\ref{fig: fresnel gain}, are used as the elements of the non-binary VR mask. Since $\nu$ depends on $\lambda_m$, the mask in the third case is influenced by the subcarrier.

\end{itemize}

\begin{figure}
    \centering
    \includesvg[width=0.70\linewidth]{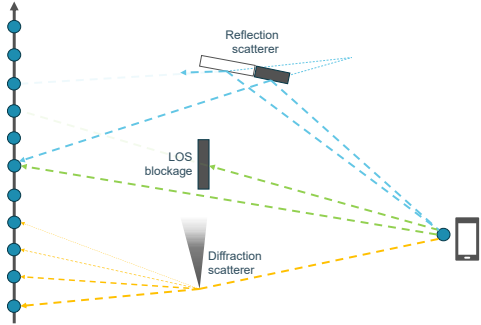}
    \caption{Illustration of SnS near-field scenario}
    \label{fig: sns channel}
\end{figure}

As a recap the near field channel with different types of spatial non-stationarity is visualized in Fig.~\ref{fig: sns channel}. 

\section{Proposed algorithm}
\label{sec: proposed algorithm}
\subsection{P-SOMP}
The proposed algorithm builds upon the existing P-SOMP~\cite{cui_channel_2022} framework for channel estimation. In this framework, the channel is transformed into a sparse representation in the polar domain as $\mathbf{h}_m = \mathbf{W}\mathbf{h}_m^\mathcal{P}$, where $\mathbf{h}^\mathcal{P}$ denotes the sparse channel representation, and $\mathbf{W}$ is the polar-domain transformation matrix, defined as:
\begin{equation}
\mathbf{W} = \left[ \mathbf{b}(\theta_1, r_{1}),\ \mathbf{b}(\theta_2, r_{2}),\ \cdots, \mathbf{b}(\theta_{\mathcal{S}}, r_{\mathcal{S}}) \right].
\label{eq: polar matrix}
\end{equation} 

The basis of the polar-domain representation consists of $\mathcal{S}$ near-field steering vectors. The optimal sampling strategy for selecting these $\mathcal{S}$ steering vectors is investigated in~\cite{cui_channel_2022}.

To perform P-SOMP, the received pilot signals are first brought together as follows: $\mathbf{y}_m = \left[ \mathbf{y}_{m,1}^T, \cdots, \mathbf{y}_{m,T}^T \right]^T$ for each subcarrier $m$, and the aggregate observation matrix is formed as $\mathbf{Y} = \left[ \mathbf{y}_1, \cdots, \mathbf{y}_M \right]$. The polar-domain transformation matrix $\mathbf{W}$ is vertically repeated $T$ times to match the dimension of the reshaped received signal.

In the P-SOMP algorithm, each iteration selects a support index, corresponding to a steering vector, based on its correlation with the current residual of the received signal. Once the support set is updated, the sparse representation is estimated by solving a least-squares problem restricted to the selected support. P-SOMP jointly exploits pilot measurements from multiple subcarriers to guide support selection, leveraging the fact that the propagation paths are shared across subcarriers. However, the sparse representations for each subcarrier are computed independently after the support selection. This procedure is repeated for $\hat{L}$ iterations, where $\hat{L}$ is the estimated sparsity level, to obtain a sparse channel representation with $\hat{L}$ non-zero components. Note that in practice, a stopping criterion is required to determine $\hat{L}$. In the simulations presented in Section~\ref{sec: simulation results}, the performance is only evaluated using $\hat{L} = 2L$.

\subsection{Interleaving HMM-based VR estimation}
The proposed algorithm interleaves a VR mask estimation into the P-SOMP framework.  After each support vector is selected and an initial estimate of the sparse representation is computed, the corresponding VR for that propagation path is estimated. This VR mask is then applied to the selected steering vector via element-wise multiplication, tailoring the steering vector to the spatially non-stationary characteristics of the channel. The sparse representation is subsequently computed again using the adapted steering vector. 

The estimation of the VR mask is performed using a two-state HMM, where each antenna is associated with one of two hidden states: \emph{in the VR} or \emph{out of the VR}. The HMM models, for each antenna, the probability of being in either state for the currently considered propagation path. The HMM is employed specifically for its ability to capture both the local observation at each antenna and the spatial correlation with neighboring antennas through its transition structure, incorporating spatial continuity and contextual information. The VR mask is then efficiently estimated by inferring the HMM with Viterbi decoding.

\subsection{Proposed VR-HMM-P-SOMP}
The full VR-HMM-P-SOMP procedure is outlined in Algorithm~\ref{algo: VR SOMP}. Steps 1–6 and 14–17 follow the original P-SOMP procedure. 
In Steps 7 and 8, an interim residual $\mathbf{R}_\Pi$ is computed, which represents the residual of the received signal without VR masking applied to the newly selected support $p^\star$. Steps 9 and 10 evaluate the contribution of this estimated path to the residual: specifically, whether it reduces or increases the residual magnitude at each antenna. Since information from multiple time slots and subcarriers is available, the antenna-wise average over these dimensions is used to form the \emph{observation} of the HMM. ($f_\text{reshape}(\cdot)$ restructures the data such that each row represents all measurements (across pilot symbols and subcarriers) observed by a single antenna.) 

In Step 11, this observation is mapped to an emission probability using a sigmoid function. This mapping ensures that, when the HMM is \emph{in the VR} state (\(s_{l,n} = 1\)), positive observations \(o_{l,n}\) yield high emission probabilities, while negative observations yield low probabilities. Conversely, in the \emph{non-VR} state (\(s_{l,n} = 0\)), negative observations correspond to high emission probabilities, and positive ones to low probabilities. The steepness of $\sigma(x)=1/(1+e^{-\mathcal{T}x})$ can be controlled with design parameter $\mathcal{T}$, and is empirically set to $20$.

In Step~12, the VR-HMM is inferred using Viterbi decoding~\cite{viterbi}. Given the known transition probabilities of the Markov model and the emission probabilities derived from the observations, the most likely sequence of hidden states is estimated. This state sequence directly defines the binary VR mask $\mathbf{s}$. In Step~13, the VR mask is applied to the current steering vector. To align with the structure of the polar domain transformation matrix $\mathbf{W}$, the mask is vertically repeated $T$ times, denoted by $\mathbf{s'}$. In the subsequent steps, the P-SOMP procedure continues, now using a support basis that has been dynamically adapted to account for spatial non-stationarity.

\begin{algorithm}
\begin{algorithmic}[1]
\caption{Proposed VR-HMM-P-SOMP}\label{algo: VR SOMP}
\STATE \textbf{Initialization:} $\mathbf{R} = \mathbf{Y},\ \mathit{\Upsilon} = \emptyset$
\FOR{$l \in \{1, 2, \cdots, \hat{L}\}$}
    \STATE Calculate the correlation matrix: $\mathbf{\Gamma} = \mathbf{W}^H \mathbf{R}$
    \STATE Select new support: $p^\star = \arg \max_p \sum_{m=1}^{M} \left| \Gamma_{p,m} \right|^2$
    \STATE Update support set: $\mathit{\Upsilon} = \mathit{\Upsilon} \cup \{p^\star\}$
    \STATE Orthogonal projection: $\hat{\mathbf{H}}_{[\mathit{\Upsilon},:]}^{\mathcal{P}} = \mathbf{W}_{[:,\mathit{\Upsilon}]}^\dagger \mathbf{Y}$

    
    \textbf{VR estimation:}
    
    \STATE Spatial Stationary path estimation: $\boldsymbol{\Pi} =  \mathbf{W}_{[:,p^\star]} \hat{\mathbf{H}}_{[p^\star, :]}^{\mathcal{P}}$
    \STATE Interim residual: $\mathbf{R}_\Pi = \mathbf{R}-\mathbf{\Pi}$
    \STATE Reshape: $\tilde{\mathbf{R}} = f_\text{reshape}\left( \mathbf{R} \right)$ and $\tilde{\mathbf{R}}_\Pi = f_\text{reshape}\left( \mathbf{R}_\Pi \right)$
    \STATE Observation: $\mathbf{o} = \frac{1}{T\cdot M} \sum_{i=1}^{T\cdot M} \left( \left| \tilde{\mathbf{R}}_{[:,i]} \right| - \left| \tilde{\mathbf{R}}_{\Pi,[:,i]} \right| \right)$
    \STATE $\mathbf{p}_{\text{emission,}s=1}= \sigma(\mathbf{o})$ and $\mathbf{p}_{\text{emission,}s=0}= \sigma(\mathbf{-o})$
    \STATE VR-mask: $\mathbf{s} = \text{Viterbi}\left( \mathbf{p}_{\text{emission,}s=1}, \mathbf{p}_{\text{emission,}s=0}, \mathbf{P}_\mathrm{trans}\right)$
    
    \hspace{0 pt}

    \STATE Mask new support vector: $\mathbf{W}_{[:,p^\star]} = \mathbf{s'} \odot \mathbf{W}_{[:,p^\star]}$
    \STATE Orthogonal projection: $\hat{\mathbf{H}}_{[\mathit{\Upsilon},:]}^{\mathcal{P}} = \mathbf{W}_{[:,\mathit{\Upsilon}]}^\dagger \mathbf{Y}$
    \STATE Update residual: $\mathbf{R} = \mathbf{Y} - \mathbf{W}_{[:,\mathit{\Upsilon}]} \hat{\mathbf{H}}_{[\mathit{\Upsilon},:]}^{\mathcal{P}}$
\ENDFOR
\STATE Final estimate: $\hat{\mathbf{H}} = \mathbf{W}_{[1:N,\mathit{\Upsilon}]} \hat{\mathbf{H}}_{[\mathit{\Upsilon},:]}^{\mathcal{P}}$ 
\end{algorithmic}
\end{algorithm}

None of the newly introduced steps (7--14) exceeds the computational complexity of Step~3, which dominates the overall complexity. Consequently, the asymptotic computational complexity of the VR-HMM-P-SOMP algorithm remains the same as that of the baseline P-SOMP algorithm, i.e., $\mathcal{O}(\hat{L}N\mathcal{S}TM)$. It should be noted, however, that this is an asymptotic analysis, and in practice, the computational workload per iteration is indeed increased due to the additional operations.

\section{Simulation results}
\label{sec: simulation results}
To evaluate the accuracy of the proposed algorithm relative to benchmark methods, a Monte Carlo simulation framework is employed. The simulation is based on the system model described in Section~\ref{sec: system model}, including the three VR mask cases. A total of $N_{\text{iter}}$ independent channel realizations are generated. Table~\ref{tab: parameters} summarizes the fixed parameters as well as those that are sampled in each iteration, outlining the complete simulation configuration.
Channel estimation accuracy is quantified using the normalized mean square error (NMSE) between the true channel matrix $\mathbf{H}$ and the estimated channel $\hat{\mathbf{H}}$, averaged over all Monte Carlo realizations.

\begin{table}
\begin{center}
\caption{Simulation configuration}
\label{tab: parameters}
\begin{tabular}{ |c | c| }
 \hline
 Parameter & Value \\
 \hline
 Number of BS antennas $N$ & 256 \\ 
 Antenna spacing $d$ & 5 mm \\
 Carrier frequency $f_c$ & 30 GHz \\
 Bandwidth $B$ & 100 MHz \\
 Number of subcarriers $M$ & 12 \\
 Number of pilot symbols $T$ & 4 \\
 Number of dominant propagation paths $L$ & 6 \\
 Channel SNR & 0~dB \\
 Distance of scatterer distribution $r$ & $\mathcal{U}(\text{7 m},\text{327 m})$ \\
 Angle of arrival distribution $\theta$ & $\mathcal{U}(\frac{-\pi}{3}, \frac{\pi}{3})$ \\
 Complex path gain distribution $g$& $\mathcal{U}(\{ g \in \mathbb{C} : |g| < 1 \})$\\
 Max diffraction angle $\theta_\text{max}$ & 0.006 \\
 Number of polar domain steering vectors $\mathcal{S}$ & 2555\\
 Temperature parameter $\mathcal{T}$ & 20 \\
 Number of paths to be detected $\hat{L}$ & 2L \\
 Transition probabilities: & \\
 -staying in VR state & $1 - \frac{1}{N}$ \\
 -changing VR state & $\frac{1}{N}$ \\
 Initial probabilities $(p(s_{l,0}=1), p(s_{l,0}=0))$ & (0.55, 0.45)\\
 Number of simulation iteration $N_{\text{iter}}$ & 500 \\
   
 \hline
\end{tabular}
\vspace{-12pt}
\end{center}
\end{table}

\begin{figure}[h]
    \centering
    \includesvg[width=0.67\linewidth]{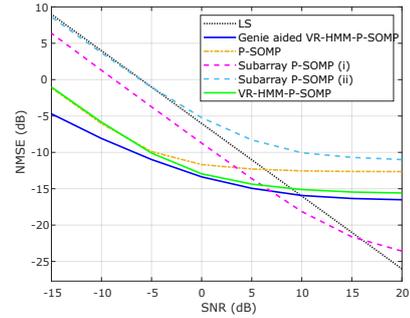}
    \caption{NMSE comparison for varying SNR.}
    \label{fig: compare snr}
    \vspace{-12pt}
\end{figure}

\begin{figure}[h]
    \centering
    \includesvg[width=0.67\linewidth]{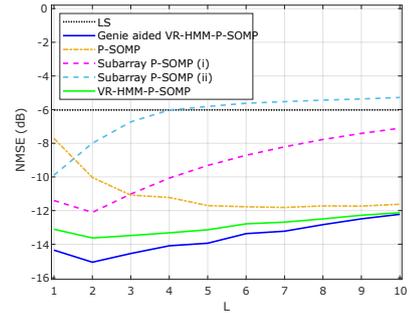}
    \caption{NMSE comparison for varying number of dominant propagation paths.}
    \label{fig: L}
\end{figure}

The proposed VR-HMM-P-SOMP algorithm is evaluated against several benchmark methods to assess its performance under varying signal-to-noise ratio (SNR) conditions. The benchmarks include least squares (LS) estimation, the original P-SOMP algorithm without VR estimation, and two variants of Subarray P-SOMP~\cite{chen_non-stationary_2024}, configured with either 8 subarrays of 32 antennas or 32 subarrays of 8 antennas. Additionally, a 'genie'-aided version of the proposed algorithm is considered, where perfect knowledge of the scatterer positions is provided. These benchmarks are compared in scenarios with different SNR. The resulting NMSE performance across SNR levels is presented in Fig.~\ref{fig: compare snr} and across sparsity levels in Fig.~\ref{fig: L}.


Fig.~\ref{fig: compare snr} shows the VR-HMM-P-SOMP algorithm consistently outperforms P-SOMP across all SNR levels, with notable NMSE gains at high SNR (e.g. $2\,\mathrm{dB}$ at $5\,\mathrm{dB}$ SNR), though the gap narrows as noise increases. Compared to subarray-based approaches, the proposed method, due to its full-array utilization, shows a clear advantage in low-SNR conditions. For instance, at -5~dB SNR, the proposed method achieves 6.5~dB NMSE gain compared to the best-performing subarray-based method. However, at higher SNR, large subarrays (32 antennas) begin to outperform the proposed method, as its NMSE curve flattens earlier. Overall, VR-HMM-P-SOMP excels in noisy environments, while the best choice at high SNR depends on subarray configuration and system design. 

Fig.~\ref{fig: L} shows that, with a larger $L$, the channel sparsity is decreasing, which results in a higher NMSE. However, the proposed method outperforms all other benchmarks, showing its robustness to the changing scattering environment.

\section{Conclusion}
This paper introduced the VR-HMM-P-SOMP algorithm to estimate channels in ELAA systems, where near-field propagation and spatial non-stationarity limit the effectiveness of traditional methods. A hybrid channel model was developed to simulate realistic ELAA conditions, combining deterministic geometry with stochastic path parameters and novel non-binary VR masks to reflect gradual power variations due to diffraction. The proposed algorithm extends the P-SOMP framework by incorporating HMM to estimate binary VR masks during sparse recovery, enabling adaptive spatial support across the array. While explicit non-binary mask estimation is left for future research, the proposed algorithm still demonstrates robust performance, particularly in low-SNR and sparse scenarios. It maintains low complexity and avoids the trade-offs of subarray-based methods, making it promising for future ELAA communication systems.

\section*{Acknowledgment}
This research is supported by the MultiX project under the Horizon Europe Research and Innovation program with Grant Agreement No. 101192521.

\balance

\bibliographystyle{IEEEtran}
\bibliography{Reference}

\begin{thebibliography}{10}
\providecommand{\url}[1]{#1}
\csname url@samestyle\endcsname
\providecommand{\newblock}{\relax}
\providecommand{\bibinfo}[2]{#2}
\providecommand{\BIBentrySTDinterwordspacing}{\spaceskip=0pt\relax}
\providecommand{\BIBentryALTinterwordstretchfactor}{4}
\providecommand{\BIBentryALTinterwordspacing}{\spaceskip=\fontdimen2\font plus
\BIBentryALTinterwordstretchfactor\fontdimen3\font minus \fontdimen4\font\relax}
\providecommand{\BIBforeignlanguage}[2]{{%
\expandafter\ifx\csname l@#1\endcsname\relax
\typeout{** WARNING: IEEEtran.bst: No hyphenation pattern has been}%
\typeout{** loaded for the language `#1'. Using the pattern for}%
\typeout{** the default language instead.}%
\else
\language=\csname l@#1\endcsname
\fi
#2}}
\providecommand{\BIBdecl}{\relax}
\BIBdecl

\bibitem{ITU-R_report_3-2516-0}
``Future technology trends of terrestrial international mobile telecommunications systems towards 2030 and beyond,'' Geneva, Zwitserland, {Rep.} {ITU}-R {M.2516-0}.

\bibitem{wang_road_2023}
C.-X. Wang, X.~You, X.~Gao, X.~Zhu, Z.~Li, C.~Zhang, H.~Wang, Y.~Huang, Y.~Chen, H.~Haas, J.~S. Thompson, E.~G. Larsson, M.~D. Renzo, W.~Tong, P.~Zhu, X.~Shen, H.~V. Poor, and L.~Hanzo, ``On the road to {6G}: Visions, requirements, key technologies, and testbeds,'' \emph{{IEEE} Communications Surveys \& Tutorials}, vol.~25, no.~2, pp. 905--974.

\bibitem{lu_tutorial_2024}
H.~Lu, Y.~Zeng, C.~You, Y.~Han, J.~Zhang, Z.~Wang, Z.~Dong, S.~Jin, C.-X. Wang, T.~Jiang, X.~You, and R.~Zhang, ``A tutorial on near-field {XL}-{MIMO} communications towards {6G},'' \emph{{IEEE} Communications Surveys \& Tutorials}, pp. 1--1.

\bibitem{zhou_spherical_2015}
Z.~Zhou, X.~Gao, J.~Fang, and Z.~Chen, ``Spherical wave channel and analysis for large linear array in {LoS} conditions,'' in \emph{2015 {IEEE} Globecom Workshops ({GC} Wkshps)}.\hskip 1em plus 0.5em minus 0.4em\relax {IEEE}, pp. 1--6.

\bibitem{liu_near-field_2024}
Y.~Liu, Z.~Wang, J.~Xu, C.~Ouyang, X.~Mu, and R.~Schober, ``Near-field communications: A tutorial review,'' pp. 1999--2049, 2023.

\bibitem{martinez2014towards}
A.~O. Martinez, E.~De~Carvalho, and J.~{\O}. Nielsen, ``Towards very large aperture massive {MIMO}: A measurement based study,'' in \emph{2014 IEEE Globecom Workshops (GC Wkshps)}.\hskip 1em plus 0.5em minus 0.4em\relax IEEE, 2014, pp. 281--286.

\bibitem{de2020non}
E.~De~Carvalho, A.~Ali, A.~Amiri, M.~Angjelichinoski, and R.~W. Heath, ``Non-stationarities in extra-large-scale massive {MIMO},'' \emph{IEEE Wireless Communications}, vol.~27, no.~4, pp. 74--80, 2020.

\bibitem{Rusek_Scaling_2013}
F.~Rusek, D.~Persson, B.~K. Lau, E.~G. Larsson, T.~L. Marzetta, O.~Edfors, and F.~Tufvesson, ``Scaling up {MIMO}: Opportunities and challenges with very large arrays,'' \emph{IEEE Signal Processing Magazine}, vol.~30, no.~1, pp. 40--60, 2013.

\bibitem{Sloane_2023-measurement}
W.~Sloane, C.~Gentile, M.~Shafi, J.~Senic, P.~A. Martin, and G.~K. Woodward, ``Measurement-based analysis of millimeter-wave channel sparsity,'' \emph{IEEE Antennas and Wireless Propagation Letters}, vol.~22, no.~4, pp. 784--788, 2023.

\bibitem{he2021wireless}
R.~He, B.~Ai, G.~Wang, M.~Yang, C.~Huang, and Z.~Zhong, ``Wireless channel sparsity: Measurement, analysis, and exploitation in estimation,'' \emph{IEEE Wireless Communications}, vol.~28, no.~4, pp. 113--119, 2021.

\bibitem{Zhang_2024-DeterministicRay}
J.~Zhang, J.~Lin, P.~Tang, W.~Fan, Z.~Yuan, X.~Liu, H.~Xu, Y.~Lyu, L.~Tian, and P.~Zhang, ``Deterministic ray tracing: A promising approach to {THz} channel modeling in {6G} deployment scenarios,'' \emph{IEEE Communications Magazine}, vol.~62, no.~2, pp. 48--54, 2024.

\bibitem{Liu2024ChannelSparsity}
X.~Liu, J.~Zhang, P.~Tang, L.~Tian, H.~Tataria, S.~Sun, and M.~Shafi, ``Channel sparsity variation and model-based analysis on 6, 26, and 105 {GHz} measurements,'' \emph{IEEE Transactions on Vehicular Technology}, vol.~73, no.~7, pp. 9387--9397, 2024.

\bibitem{bajwa_compressed_2010}
W.~U. Bajwa, J.~Haupt, A.~M. Sayeed, and R.~Nowak, ``Compressed channel sensing: A new approach to estimating sparse multipath channels,'' \emph{Proceedings of the {IEEE}}, vol.~98, no.~6, pp. 1058--1076.

\bibitem{berger_application_2010}
C.~R. Berger, Z.~Wang, J.~Huang, and S.~Zhou, ``Application of compressive sensing to sparse channel estimation,'' \emph{{IEEE} Communications Magazine}, vol.~48, no.~11, pp. 164--174.

\bibitem{mansoor_massive-mimo_2017}
B.~Mansoor, S.~Nawaz, and S.~Gulfam, ``Massive-{MIMO} sparse uplink channel estimation using implicit training and compressed sensing,'' \emph{Applied Sciences}, vol.~7, no.~1, p.~63.

\bibitem{Liu_2016_joint-burst}
A.~Liu, V.~Lau, and W.~Dai, ``Joint burst {LASSO} for sparse channel estimation in multi-user massive {MIMO},'' in \emph{2016 IEEE International Conference on Communications (ICC)}, 2016, pp. 1--6.

\bibitem{schniter2014channel}
P.~Schniter and A.~Sayeed, ``Channel estimation and precoder design for millimeter-wave communications: The sparse way,'' in \emph{2014 48th Asilomar Conference on Signals, Systems and Computers}, 2014, pp. 273--277.

\bibitem{destino2015leveraging}
G.~Destino, M.~Juntti, and S.~Nagaraj, ``Leveraging sparsity into massive {MIMO} channel estimation with the adaptive-{LASSO},'' in \emph{2015 IEEE Global Conference on Signal and Information Processing (GlobalSIP)}, 2015, pp. 166--170.

\bibitem{xu_joint_2024}
W.~Xu, A.~Liu, and M.-j. Zhao, ``Joint visibility region detection and channel estimation for {XL}-{MIMO} systems via alternating {MAP}.''

\bibitem{yu2023channel}
X.~Yu, W.~Shen, R.~Zhang, C.~Xing, and T.~Q.~S. Quek, ``Channel estimation for {XL}-{RIS}-aided millimeter-wave systems,'' \emph{IEEE Transactions on Communications}, vol.~71, no.~9, pp. 5519--5533, 2023.

\bibitem{chen2022hierarchical}
J.~Chen, P.~Zhang, N.~Ma, and X.~Xu, ``Hierarchical-block sparse bayesian learning for spatial non-stationary massive {MIMO} channel estimation,'' \emph{IEEE Wireless Communications Letters}, vol.~11, no.~5, pp. 888--892, 2022.

\bibitem{cheng2019adaptive}
X.~Cheng, K.~Xu, J.~Sun, and S.~Li, ``Adaptive grouping sparse bayesian learning for channel estimation in non-stationary uplink massive {MIMO} systems,'' \emph{IEEE Transactions on Wireless Communications}, vol.~18, no.~8, pp. 4184--4198, 2019.

\bibitem{tropp_greed_2004}
J.~Tropp, ``Greed is good: Algorithmic results for sparse approximation,'' \emph{{IEEE} Transactions on Information Theory}, vol.~50, no.~10, pp. 2231--2242.

\bibitem{cui_channel_2022}
M.~Cui and L.~Dai, ``Channel estimation for extremely large-scale {MIMO}: Far-field or near-field?'' \emph{{IEEE} Transactions on Communications}, vol.~70, no.~4, pp. 2663--2677.

\bibitem{yuan_spatial_2023}
Z.~Yuan, J.~Zhang, Y.~Ji, G.~F. Pedersen, and W.~Fan, ``Spatial non-stationary near-field channel modeling and validation for massive {MIMO} systems,'' \emph{{IEEE} Transactions on Antennas and Propagation}, vol.~71, no.~1, pp. 921--933.

\bibitem{han_channel_2020}
Y.~Han, S.~Jin, C.-K. Wen, and X.~Ma, ``Channel estimation for extremely large-scale massive {MIMO} systems,'' \emph{{IEEE} Wireless Communications Letters}, vol.~9, no.~5, pp. 633--637.

\bibitem{chen_non-stationary_2024}
Y.~Chen and L.~Dai, ``Non-stationary channel estimation for extremely large-scale {MIMO},'' \emph{{IEEE} Transactions on Wireless Communications}, vol.~23, no.~7, pp. 7683--7697.

\bibitem{tang_joint_2024}
A.~Tang, J.-B. Wang, Y.~Pan, W.~Zhang, X.~Zhang, Y.~Chen, H.~Yu, and R.~C. De~Lamare, ``Joint visibility region and channel estimation for extremely large-scale {MIMO} systems,'' \emph{{IEEE} Transactions on Communications}, vol.~72, no.~10, pp. 6087--6101.

\bibitem{sherman_properties_1962}
J.~Sherman, ``Properties of focused apertures in the fresnel region,'' \emph{{IRE} Transactions on Antennas and Propagation}, vol.~10, no.~4, pp. 399--408.

\bibitem{ITU_r_p.526}
``Propagation by diffraction,'' Geneva, Zwitserland, 2019, {Rec.} {ITU}-{R} {P.}526–15.

\bibitem{viterbi}
G.~Forney, ``The viterbi algorithm,'' \emph{Proceedings of the IEEE}, vol.~61, no.~3, pp. 268--278, 1973.

\end{thebibliography}

\end{document}